\documentclass[a4paper,twoside,runningheads]{llncs}

\usepackage[T1]{fontenc}
\usepackage{times}
\usepackage{hyperref}
\usepackage{amsmath}
\usepackage{amssymb}
\usepackage{amsthm}
\usepackage[dvipsnames]{xcolor}
\usepackage{pdfpages}

\allowdisplaybreaks
\newif\ifarxiv
\arxivtrue
\newif\ifreview
\reviewfalse

\ifarxiv
\reviewfalse
\pdfoutput=1 
\else
\fi

\usepackage{stmaryrd}
\usepackage{laws}

\newcommand{\draftonly}[1]{}

\newcommand{\fun}{\mathrel{\rightarrow}}
\newcommand{\refsto}{\mathrel{\succeq}}
\newcommand{\refines}{\mathrel{\preceq}}
\newcommand{\nondet}{\mathbin{\vee}}
\newcommand{\Nondet}{\mathbin{\textstyle\bigvee}}
\newcommand{\meet}{\mathbin{\wedge}}
\newcommand{\Meet}{\mathop{\bigwedge}}
\newcommand{\Seq}{\mathbin{;}}
\newcommand{\together}{\mathbin{\Cap}}

\newcommand{\Nil}{\boldsymbol{\tau}}
\newcommand{\cgd}[1]{\mathop{\tau}#1}
\newcommand{\pstepd}{\pi}
\newcommand{\estepd}{\epsilon}

\newcommand{\cstepd}{\boldsymbol{\alpha}}
\newcommand{\cstep}[1]{\mathop{\alpha}#1}
\newcommand{\cpstepd}{\boldsymbol{\pstepd}}
\newcommand{\cpstep}[1]{\mathop{\pstepd}#1}
\newcommand{\cestepd}{\boldsymbol{\estepd}}
\newcommand{\cestep}[1]{\mathop{\estepd}#1}
\newcommand{\Abort}{\top}
\newcommand{\Magic}{\bot}

\newcommand{\ptran}[2]{#1 \xrightarrow{\pi} #2}
\newcommand{\etran}[2]{#1 \xrightarrow{\epsilon} #2}
\newcommand{\ptranssp}{\ptran{\sigma}{\sigma'}}
\newcommand{\etranssp}{\etran{\sigma}{\sigma'}}
\newcommand{\Command}{\mathcal{C}}

\newcommand{\kw}[1]{\mathbf{\color{black}#1}}

\newcommand{\Do}{\mathbin{\kw{do}}}
\newcommand{\Else}{\mathbin{\kw{else}}}
\newcommand{\Evolve}{\mathop{\kw{evolve}}}
\newcommand{\lblot}{\langle\mkern -3.5mu{|}}
\newcommand{\rblot}{|\mkern -3.5mu{\rangle}}
\newcommand{\Ex}[1]{#1}
\newcommand{\LEx}[1]{#1}
\newcommand{\Expr}[2]{\Ex{\lblot#1\rblot_{#2}}}

\newcommand{\Eval}[2]{\Ex{\llbracket#1\rrbracket}_{#2}}

\newcommand{\Const}[1]{\Ex{#1}}
\newcommand{\Unary}[2]{\Ex{\mathop{#1}#2}}
\newcommand{\Binary}[3]{\Ex{#1\mathbin{#2}#3}}
\newcommand{\Deref}[1]{\Ex{{*}\LEx{#1}}}
\newcommand{\Variable}[1]{\LEx{#1}}
\newcommand{\Base}[1]{\LEx{#1}}
\newcommand{\Array}[2]{\LEx{#1[\Ex{#2}]}}
\newcommand{\Indexed}[2]{\LEx{#1\left[\Ex{#2}\right]}}
\newcommand{\UnaryOp}{\mathop{\ominus}}
\newcommand{\BinaryOp}{\mathop{\oplus}}
\newcommand{\Fi}{\mathop{\kw{fi}}}
\newcommand{\Guarantee}{\mathop{\kw{guar}_{\boldsymbol{\pi}}}}

\newcommand{\If}{\mathop{\kw{if}}}

\newcommand {\inv}{\mathop{\kw{inv}}}
\newcommand{\Od}{\mathbin{\kw{od}}{}}

\newcommand{\Rely}{\mathop{\kw{rely}}}

\newcommand{\Then}{\mathbin{\kw{then}}}

\newcommand{\Let}{\mathop{\kw{let}}}
\newcommand{\In}{\mathbin{\kw{in}}}
\newcommand{\Call}{\mathop{\kw{call}}}
\newcommand{\While}{\mathop{\kw{while}}}

\newcommand{\UndefinedValue}{\lightning}

\newcommand{\GFP}{\nu}
\newcommand{\LFP}{\mu}

\newcommand{\lBrace}{\{\mkern -4.5mu{|}}
\newcommand{\rBrace}{|\mkern -4.5mu{\}}}
\newcommand{\Pre}[1]{\lBrace#1\rBrace}

\newcommand{\Fin}[1]{#1^{\star}}
\newcommand{\Finrel}[1]{#1^{*}}
\newcommand{\Om}[1]{#1^{\omega}}

\newcommand{\wfr}{\mathrel{<\hspace{-5pt}<}}
\newcommand{\wfreq}{\mathrel{\leq\hspace{-5pt}\leq}}
\newcommand{\WFrelation}[2]{#1 \wfr #2}

\newcommand{\GtWFEval}[2]{\Set{#1 \wfr #2}}

\makeatletter
\newcommand{\Comprehension}[3]{\{\ifx\@empty#3\else#3 \fi \ifx\@empty#1\else\mid #1\fi \mathrel{.} #2 \}}
\makeatother

\newcommand{\atomicrel}[1]{\langle#1\rangle}

\newcommand{\evolve}[1]{\Evolve #1}
\newcommand{\Frame}[2]{\ifx\@empty#1\else\LEx{#1}:\fi#2}
\newcommand{\guar}[1]{\Guarantee #1}
\newcommand{\Idle}{\kw{idle}}
\newcommand{\opt}[1]{\mathop{\kw{opt}}#1}
\newcommand{\Post}[1]{\Spec{}{}{#1}}
\newcommand{\rely}[1]{\Rely #1}

\newcommand{\Comment}[1]{\mbox{{\color{purple}--- #1}}}
\makeatletter
\def\Spec{\@ifnextchar*{\@Spec}{\@@Spec}}
\def\@Spec*#1#2#3{\ifx\@empty#1\else#1:\fi
   \llparenthesis{#2}\ifx\@empty#2\else,~\fi#3\rrparenthesis}
\def\@@Spec#1#2#3{\ifx\@empty#1\else
   \begin{array}{@{}l@{}}#1\end{array}:\fi%
   \left(\hspace*{-2pt}\left|{\begin{array}{@{}l@{}}#2\end{array}}\ifx\@empty#2\else~,~~\fi
   \begin{array}{@{}l@{}}#3\end{array}\right|\hspace*{-2pt}\right)}
\makeatother
\newcommand{\Term}{\kw{term}}
\newcommand{\Fair}{\kw{fair}}

\newcommand{\rename}[2]{[#1\backslash#2]}

\newenvironment{RelatedWork}{\paragraph{Related work.}}{\hfill$\Box$}

\newcommand{\EqEval}[2]{\Set{#1 = #2}}

\newcommand{\Mid}{\frac{lo + hi}{2}}

\makeatletter
\def\Set{\@ifnextchar*{\@Set}{\@@Set}}
\def\@Set*#1{{\color{blue}\left\llcorner\begin{array}{l}#1\end{array}\right\lrcorner}}
\def\@@Set#1{{\color{blue}\llcorner#1\lrcorner}}
\def\Rel{\@ifnextchar*{\@Rel}{\@@Rel}}
\def\@Rel*#1{{\color{blue}\left\ulcorner\begin{array}[t]{l}#1\end{array}\right\urcorner}}
\def\@@Rel#1{{\color{blue}\ulcorner#1\urcorner}} 
\def\RelA{\@ifnextchar*{\@RelA}{\@@RelA}}
\def\@RelA*#1{{\color{purple}\left\lceil\BB\left\lceil\begin{array}{l}#1\end{array}\right\rceil\BB\right\rceil}}
\def\@@RelA#1{{\color{purple}\lceil\BB\lceil#1\rceil\BB\rceil}}
\def\Ratomicrel{\@ifnextchar*{\@Ratomicrel}{\@@Ratomicrel}}
\def\@Ratomicrel*#1{\left\langle\Rel*{#1}\right\rangle}
\def\@@Ratomicrel#1{\atomicrel{\Rel{#1}}}
\makeatother
\newcommand{\Spre}[1]{\Pre{\Set{#1}}}
\makeatletter
\def\Rspec{\@ifnextchar*{\@Rspec}{\@@Rspec}}
\def\@Rspec*#1#2#3{\Spec{#1}{#2}{\Rel*{#3}}}
\def\@@Rspec#1#2#3{\Spec{#1}{#2}{\Rel{#3}}}
\makeatother

\newcommand{\Rrely}[1]{\rely{\Rel{#1}}}

\newcommand{\Revolve}[1]{\evolve{\Rel{#1}}}
\newcommand{\Sinv}[1]{\inv{\Set{#1}}}

\newcommand{\defs}{\mathrel{\widehat=}}
\renewcommand{\implies}{\mathrel{\Rightarrow}}

\newcommand{\bool}{\mathbb{B}}
\newcommand{\True}{\mathsf{true}}
\newcommand{\False}{\mathsf{false}}
\newcommand{\dom}{\mathop{\kw{dom}}}
\makeatletter
\def\comp@sym{\raise 0.6ex\hbox{\small\oalign{\hfil%
        $\scriptscriptstyle\mathrm{o}$\hfil%
        \cr\hfil$\scriptscriptstyle\mathrm{9}$\hfil}}}
\newcommand{\semi}{\mathrel{\comp@sym}}
\makeatother
\newcommand{\compl}[1]{\overline{#1}}
\newcommand{\id}[1]{{\textstyle\mathsf{id}}_{#1}}
\newcommand{\universalrel}{\mathsf{univ}}
\newcommand{\dres}{\mathbin{\vartriangleleft}}
\newcommand{\rres}{\mathbin{\vartriangleright}}

\newcommand{\postr}[1]{\overrightarrow{#1}}
\newcommand{\inter}{\mathbin{\cap}}

\newcommand{\union}{\mathbin{\cup}}
\newcommand{\Union}{\mathop{\bigcup}}
\newcommand{\spot}{{.~}}

\newcommand{\Triple}[3]{\{#1\}\; #2\; \{#3\}}

\newcommand{\Establish}[4]{\Triple{#1}{\rely{#2} \together #3}{#4}}

\newcommand{\EstablishExpr}[5]{\Establish{#1}{#2}{\Expr{#3}{#4}}{#5}}

\makeatletter
\newcommand{\ChainRel}[1]{\crcr \noalign{\penalty\interdisplaylinepenalty}
  \hspace*{-1em}#1 &
  \@ifnextchar*{\@ChainRelCommment}{}}
\newcommand{\Why}[1]{\mbox{{\color{blue}\hspace*{0.5em}#1}}}
\def\@ChainRelCommment*[#1]{\Why{#1}
  \crcr & 
  }
\newcommand{\StartRef}[1]{\hspace*{-1.5em} \ref{#1}) \refsto
  \@ifnextchar[{\@StartRefCommment}{}}
\def\@StartRefCommment[#1]{\mbox{#1}
  \crcr \noalign{\penalty\interdisplaylinepenalty}}
\makeatother
\newcommand{\Implies}{\ChainRel{\implies}}

\newcommand{\entails}{\Rrightarrow}
\newcommand{\Entails}{\ChainRel{\entails}}
\newcommand{\Equiv}{\ChainRel{\equiv}}
\newcommand{\Refsto}{\ChainRel{\refsto}}
\newcommand{\Equals}{\ChainRel{=}}

\makeatletter
\def\@setmcodes#1#2#3{{\count0=#1 \count1=#3
  \loop \global\mathcode\count0=\count1 \ifnum \count0<#2
  \advance\count0 by1 \advance\count1 by1 \repeat}}

\DeclareSymbolFont{italic}{OT1}{\rmdefault}{m}{it}
\let\mathit\undefined
\DeclareSymbolFontAlphabet{\mathit}{italic}
\edef\@tempa{\hexnumber@\symitalic}
\@setmcodes{`A}{`Z}{"7\@tempa41}
\@setmcodes{`a}{`z}{"7\@tempa61}
\makeatother

\makeatletter
\newcommand{\InfRule}[3]{
    { \boxed{#1} 
      \if\@empty#2\strut\displaystyle \begin{array}{c}#3 \end{array}\else
          \frac {\strut\displaystyle \begin{array}{c}#2 \end{array}} 
                  {\strut\displaystyle \begin{array}{c}#3 \end{array}}\fi
      \hfil }
}
\makeatother

\newcommand\numberthis{\addtocounter{equation}{1}\tag{\theequation}}

\definecolor{CJ}{rgb}{1,1,0.9}
\definecolor{IH}{rgb}{1,0.9,1}
\definecolor{LM}{rgb}{0.9,1,1}

\newcounter{hours}
\newcounter{minutes}
\newcommand{\printtime}{%
  \ifthenelse{\value{hours}<10}{0}{}\thehours:%
  \ifthenelse{\value{minutes}<10}{0}{}\theminutes}

\newbox{\MyDate}
\savebox{\MyDate}{%
\ifarxiv \scriptsize(\today\ \printtime)\fi%
}

\title{Reasoning about concurrent loops and recursion with rely-guarantee rules}
\titlerunning{Reasoning about concurrent loops and recursion \usebox{\MyDate}}
\author{
Ian J. Hayes\inst{1}\orcidID{0000-0003-3649-392X}
\and
Larissa A. Meinicke\inst{1}\orcidID{0000-0002-5272-820X}
\and
Cliff B. Jones\inst{2}\orcidID{0000-0002-0038-6623}
\titlerunning{Rely-guarantee reasoning about loops and recursion\usebox{\MyDate}} 
\institute{School of Electrical Engineering and Computer Science, \\ 
The University of Queensland, Brisbane, Queensland 4072, Australia \and
School of Computing, Newcastle University, UK
}
}

\authorrunning{I. J. Hayes,  L. A. Meincke and C. B. Jones \usebox{\MyDate}} 
\institute{
School of Electrical Engineering and Computer Science, \\ 
The University of Queensland, Brisbane, Queensland 4072, Australia
}

\begin{document}

\ifreview
\let\origthepage=\thepage
\makeatletter
\renewcommand{\thepage}{\@arabic\c@page-R}
\makeatother
\clearpage
\let\thepage=\origthepage
\setcounter{page}{1}
\setcounter{section}{0}
\write128{Start of main material: \the\ReadonlyShipoutCounter.}
\fi

\maketitle

\begin{abstract}
The objective of this paper is to present general, mechanically verified, refinement rules for reasoning about recursive programs and while loops in the context of concurrency.
We make use of the rely-guarantee approach to concurrency 
that facilitates reasoning about interference from concurrent threads in a compositional manner.
Recursive programs can be defined as fixed points over a lattice of commands
and hence we develop laws for reasoning about fixed points. 
Loops can be defined in terms of fixed points
and hence the laws for recursion can be applied to develop laws for loops.
Unlike many approaches to concurrency,
we do not assume that expression evaluation is atomic.
\end{abstract}

\section{Introduction}\labelsect{introduction}

For sequential programs, the Hoare-style verification rule for a while loop \cite{Hoare69a,Gries81}, 
\begin{align}
\InfRule
{\While}
{{\Triple{p \land b \land z = z_0}{c}{p \land z \wfr z_0}}}
{\Triple{p}{\While b \Do c \Od}{p \land \neg b}} \labelprop{Hoare-loop}
\end{align}
makes use of a loop invariant, $p$,
that is assumed to hold initially and is maintained by the body of the loop if the loop guard $b$ holds.
To show termination, a variant expression $z$ is required to 
strictly decrease on each iteration according to a well-founded relation $\wfr$.
Rule \refprop{Hoare-loop} is not valid if there are concurrent threads modifying shared variables
because interference can
(i) invalidate the invariant $p$ or
(ii) increase the variant expression $z$ or
(iii) interfere with the evaluation of $b$ 
in such a way that $b$ may not hold at the start of the execution of the loop body or
(iv) prevent $\lnot b$ holding on exit from the loop.

Finding compositional rules that cope with interference proved challenging (see \cite{DeRoever01});
Rely/Guarantee (RG) conditions offer a compositional way to develop shared-variable concurrent programs \cite{Jones81d,Jones83a,Jones83b};
initially the inference rules were presented in the keyword style of VDM \cite{Jones90a};
a series of contributions \cite{AFfGRGRACP,Concurrent_Ref_Alg-AFP,FM2016atomicSteps,FMJournalAtomicSteps,2024RAMiCSrestructuring}
succeeded in recasting the RG approach in an algebraic style. 

When designing a thread $T$,
it is assumed that all atomic state-to-state transitions made by its environment (i.e.\ concurrently running threads)
satisfy a rely condition $r$, a relation between program states.
To complement this, the implementation of thread $T$ must ensure that all program transitions it makes satisfy a guarantee  relation $g$,
where $g$ must imply the rely condition of every thread running in parallel with $T$.
\reffig{rely-guar} gives an example of an Aczel trace of program and environment transitions \cite{Aczel83,DeRoever01,DaSMfaWSLwC}.
The intricate question of which transitions are to be viewed as atomic is addressed in \refsect{language}.

\begin{figure}[t]
\begin{center}
\input{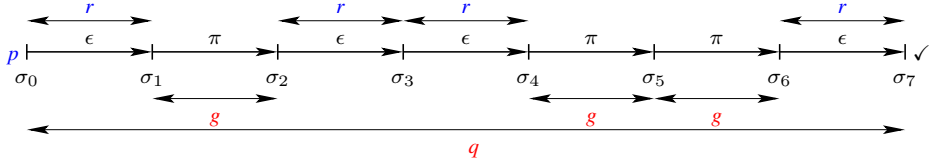}
\end{center}\vspace{-3ex}
\caption{An Aczel trace satisfying a specification with precondition $p$, postcondition $q$, rely condition $r$ and guarantee condition $g$.
If the initial state, $\sigma_0$, is in $p$ and all environment transitions ($\estepd$) satisfy $r$,
then all program transitions ($\pstepd$) must satisfy $g$ and 
the initial state $\sigma_0$ must be related to the final state $\sigma_7$ by the postcondition $q$, also a relation between states.
}\labelfig{rely-guar}
\vspace{-2ex}
\end{figure}

This paper develops laws for reasoning about recursion (\refsect{recursion}) and while loops (\refsect{loop}) in a rely-guarantee style.
The proof of termination of a recursion makes use of well-founded induction with a variant function and a well-founded relation;
however, the standard approach of requiring the variant to be reduced by the function body before any recursive calls are made 
does not deal with interference from other threads reducing the variant in a manner that precludes the function body reducing it.
Hence we present an early termination version of the law that allows the function body to either reduce the variant before recursive calls
or establish a condition such that a call on the function will not itself make any recursive calls
(i.e.\ it will only use the base case(s) of the function).
Because a while loop is defined as a recursive function, 
our law for while loops is proven in terms of the law for recursion;
it too is an early termination version.
An example of developing a recursive program that allows concurrent threads to share variables is given in \refsect{parallel-search};
the use of a while loop that can terminate prematurely is illustrated in \refsect{ex-early-termin}.

\begin{RelatedWork}
It is interesting to compare Concurrent Separation Logic (CSL) with the Rely-Guarantee ideas behind this paper.
The aim of Reynolds' Separation Logic is to offer a way to reason about ownership of heap addresses \cite{SepLogicReynolds};
O'Hearn showed its applicability to concurrent heap programs in \cite{OHearn07}.
In contrast, the early RG publications were applied to stack variables although there is no reason to avoid programs that employ heaps.
Subsequently a variety of CSLs have been applied to reasoning about ownership of other resources including locks.
(A useful historical account of the origins and developments of CSLs is \cite{BrookesOHearn-16}.)
One attempt to classify the application areas of the two approaches is stated by O'Hearn in \cite{OHearn07} to be that CSL supports reasoning about race freedom whereas RG allows proofs about ``racy'' programs.
Whilst this is superficially true,
it ignores development by reifying abstractions
and it is important to remember the initial emphasis of RG research was precisely to provide a compositional development method for shared-variable concurrent programs.
For example \cite{PVEaCfC} addresses Asynchronous Communication Mechanisms and presents a development of Hugo Simpson's ``four-slot implementation''
whose essence is to avoid races on the slots;
the RG development identifies the conditions for race freedom at an abstract level.
There are two important combinations of CSL and RG rules: RGSep \cite{Vafeiadis07} and SAGL \cite{FengSAGL-07};
an alternative approach to viewing separation as an abstraction is offered in \cite{JonesYatapanage-15}.
Bornat also looked at employing combinations of the approaches on proofs about Simpson's algorithm in \cite{BornatAmjad10,BornatAmjad13} 
but \cite{BornatAmjad13} does not really use CSL.

Most RG approaches 
\cite{Jones81d,Jones83a,Jones83b,Stolen90,stolen1991method,XuRoeverHe97,Dingel02,PrensaNieto03,Wickerson10-TR,DBLP:conf/esop/WickersonDP10,SchellhornTEPR14,Sanan21}, 
employ laws that only apply if expression evaluation is atomic,
which does not reflect the reality of expression evaluation in programming language implementations.
Such laws can only be used with expressions that have a single critical reference 
\cite[Definition 2.7, p.27]{BenAri2005} \cite[Definition 2.4]{Andrews1991};
but the validity of their application in this case cannot be verified within their theories
precisely because of the atomicity assumption in their semantics;
hence they break the formal chain of reasoning. 
An exception is the work of Coleman and Jones \cite{CoJo07,Coleman08} 
who develop a fine-grained operational semantics that does not assume expression evaluation is atomic,
however, the rules they develop are restricted to simple special cases of our rules.
We assume neither expression evaluations nor assignment commands are executed atomically
(see \cite{2026MultiReferenceExpressions}).
\end{RelatedWork}

\section{Wide-spectrum language}\labelsect{language}

This section introduces our wide-spectrum language
that incorporates both specification and programming language commands \cite{CIP}.
Our approach follows that used for the sequential refinement calculus \cite{BackWright98,Morgan94}
but extends it with concurrency, which requires a richer semantics based on Aczel traces \cite{Aczel83,DeRoever01,DaSMfaWSLwC}.
The underlying theory is based on our earlier research 
\cite{AFfGRGRACP,Concurrent_Ref_Alg-AFP,FM2016atomicSteps,FMJournalAtomicSteps,2024RAMiCSrestructuring}
and all lemmas and theorems have been proven in our corresponding Isabelle/HOL theories.
\ifreview\marginpar{\color{purple}For reviewing purposes, a pdf of the Isabelle/HOL theories for expressions and commands are included after the Appendix.}\fi

\paragraph{Naming and syntactic precedence conventions.}

We use 
$\Sigma$ for the set of program states;
$\sigma$ for program states (i.e.\ $\sigma \in \Sigma$);
$c$ and $d$ for commands;
$p$ for sets of program states;
$g$, $q$ and $r$ for binary relations between program states;
$u$, $v$ and $w$ for program variables;
$e$ for expressions;
$b$ for boolean expressions, where $\bool$ is the set of booleans;
and 
$k$ and $\kappa$ for values.
Subscripted versions of the above names follow the same convention.
Unary operations and function application have higher syntactic precedence than binary operations.
For binary operations, non-deterministic choice ($\nondet$) has the lowest precedence, 
and sequential composition ($\Seq$) has the highest precedence,
except framing ($\Frame{V}{c}$) is higher.
We use parentheses to resolve all other syntactic ambiguities.

\subsubsection{Values, left values and states.}\labelsect{state}

Our theory treats values, left values (variables and indexed array elements) and program states abstractly.
For values ($\in Value$) the only constraint is that $Value$ contains the distinct boolean values $\True$ and $\False$
(so that we can handle boolean guards in conditionals and loops)
and the undefined value $\UndefinedValue$ (so we can handle divide-by-zero, out of bounds array indexing, etc.).
Left values (l-values for short) consist of base variable names ($v \in Identifier$) and 
indexed array elements ($\LEx{Indexed\,A\,\Ex{i}}$), where $A$ is an l-value and $i$ is an index value,
similar to the approach used by Dingel \cite{Dingel02}.
Importantly, evaluation of an l-value for an indexed array element is done before dereferencing its value,
so that atomicity is only required for accessing individual array elements, not the whole array.
\begin{align}
  LValue \defs Base\,Identifier \mid Indexed\,LValue\,Value  \labeldef{LValue}
\end{align}
We abbreviate ``$\LEx{Base\,v}$'' to ``$\Base{v}$'', 
and ``$\LEx{Indexed\,A\,\Ex{i}}$'' to ``$\Indexed{A}{i}$''.
L-values could be extended to handle a heap but we do not consider that in this paper.%
\footnote{Because a heap corresponds to an anonymous array with indices being pointer values,
the theory for handling arrays could be used to handle a heap.}
The set of possible program states, $\Sigma$, is a mapping from l-values to values:
\(
  \Sigma \defs LValue \fun Value . 
\)

\subsubsection{Command lattice.}\labelsect{lattice}

Our language consists of a complete distributive lattice of commands, $\mathcal{C}$,
with partial order, $c \refsto d$, meaning command $c$ is refined (or implemented) by command $d$,
so that non-deterministic choice ($c \nondet d$) is the lattice join
and strong conjunction ($c \meet d$) is the lattice meet.
The lattice is complete so that, for a set of commands $C$, 
non-deterministic choice, $\Nondet C$, 
and strong conjunction, $\Meet C$, are 
are defined as the least upper bound
and greatest lower bound, respectively,
and the bottom and top elements of the lattice are $\bot$ and $\top$, respectively.
The lattice operators 
are considered part of our wide spectrum language.
Given a complete lattice we can define least ($\mu$) and greatest ($\nu$) fixed points of monotone functions.%
\footnote{In other works on the refinement calculus \cite{BackWright98,BackWright99}, the dual lattice is used
(i.e.\ with the reverse ordering), so greatest fixed points in our theory correspond to least fixed points in theirs.}
Programming language functions are defined via \emph{parametrised commands}, that is, they are of type, $Value \fun \Command$.

\subsubsection{Aczel trace semantics and primitive operations.}\labelsect{semantics}

The semantic model for the lattice of commands consists of sets of Aczel traces \cite{Aczel83,DeRoever01,DaSMfaWSLwC}.
An Aczel trace consists of a possibly infinite sequence of state-to-state transitions 
each of which is labeled as either a program ($\pstepd$) or environment ($\estepd$) transition, as in \reffig{rely-guar}.
A trace may be either terminating, aborting or incomplete.
Sets of traces corresponding to commands are 
prefix closed (i.e.\ they contain all incomplete prefix traces of a trace)
and abort closed (i.e.\ they contain all possible extensions of an aborting trace).
The language includes the following three binary operators:
\begin{description}
\item[$c \Seq d$,]
sequential composition, that is associative, has the null command, $\Nil$, as its neutral element,
and distributes arbitrary non-deterministic choices in its left argument \refax{Nondet-distrib-seq-right} 
and non-empty non-deterministic choices in its right argument \refax{Nondet-distrib-seq-left}.
A trace of $c \Seq d$ consists of the gluing concatenation of a terminating trace of $c$ and a trace of $d$,
or an incomplete (including infinite) or aborting trace of $c$.
A \emph{gluing concatenation} of a terminating trace $tr_1$ with a trace $tr_2$ 
requires that the final state of $tr_1$ equals the initial state of $tr_2$ 
and these two states become a single state in the concatenation.
\item[$c \parallel d$,]
parallel composition, that is associative, commutative, 
and distributes non-empty non-deterministic choices in both its arguments \refax{Nondet-distrib-par} 
because $\parallel$ is commutative.
Because each trace records \emph{both} the behaviour of the program \emph{and} its environment, 
parallel composition is synchronous \cite{Milner83,FM2016atomicSteps,FMJournalAtomicSteps}, and 
a trace of $c \parallel d$ consists of a matching of a trace of $c$ and a trace of $d$,
where a program transition $\ptranssp$ of $c$ matches an environment transition $\etranssp$ of $d$ to give $\ptranssp$ for $c \parallel d$ 
(or vice versa)
and an environment transition $\etranssp$ for both $c$ and $d$ matches to give the same environment transition for $c \parallel d$.
Note that a program transition of $c$ cannot match a program transition of $d$,
leading to the interleaving of their program transitions.
If either $c$ or $d$ aborts, so does $c \parallel d$.
A terminating trace of $c \parallel d$ corresponds to matching terminating traces of $c$ and $d$.
\item[$c \together d$,]
weak conjunction, that is associative, commutative, idempotent, 
and distributes non-empty non-deterministic choices in both its arguments \refax{Nondet-distrib-conj} 
because $\together$ is commutative.
A trace of $c \together d$ is either a trace of both $c$ and $d$, 
or it is an aborting trace of $c$ that equals an incomplete trace of $d$ up to the point of abort of $c$ (or vice versa).
Weak conjunction ($c \together d$) behaves as both $c$ and $d$ unless either $c$ or $d$ aborts, 
in which case the weak conjunction aborts.
Termination of $c \together d$ corresponds to termination of both $c$ and $d$, that is,
both $c$ and $d$ execute the same terminating trace.
\end{description}
\begin{align}
  \Nondet_{c \in C} (c \Seq d) & = (\Nondet_{c \in C} c) \Seq d \labelax{Nondet-distrib-seq-right} \\
  \Nondet_{c \in C} (d \Seq c) & = d \Seq (\Nondet_{c \in C} c)  & \mbox{if } C \neq \emptyset \labelax{Nondet-distrib-seq-left} \\
  \Nondet_{c \in C} (c \parallel d) & = (\Nondet_{c \in C} c) \parallel d  & \mbox{if } C \neq \emptyset \labelax{Nondet-distrib-par} \\
  \Nondet_{c \in C} (c \together d) & = (\Nondet_{c \in C} c) \together d  & \mbox{if } C \neq \emptyset \labelax{Nondet-distrib-conj} 
\end{align}

\subsubsection{Primitive commands.}\labelsect{primitives}

A command is \emph{infeasible} in a state, $\sigma$, if it cannot make any transition from $\sigma$
and it can neither terminate nor abort in state $\sigma$.
The language includes five primitive commands:
\begin{description}
\item[$\Magic$~]
is the command that is everywhere infeasible. It is the least command in our lattice;
it satisfies $\Magic \Seq c = \Magic$, for any command $c$.
\item[$\cgd{p}$]
is an instantaneous test that the current state $\sigma$ is in the set of states $p$:
if $\sigma \in p$, $\cgd{p}$ terminates, otherwise it is infeasible -- 
note that there is no transition taken by an instantaneous test 
(in the trace semantics, its trace is tagged as terminated);
\item[$\cpstep{r}$]
is an atomic program command 
that may perform a program transition $\ptranssp$ if $(\sigma,\sigma') \in r$ and then terminate,
otherwise it is infeasible if $\sigma \not\in \dom r$;
\item[$\cestep{r}$]
is an atomic environment command 
that may perform an environment transition $\etranssp$ if $(\sigma,\sigma') \in r$ and then terminate,
otherwise it is infeasible, 
and
\item[$\Abort$~]
is Dijkstra's abort command \cite{Dijkstra75,Dijkstra76} that can do any behaviour whatsoever.
It is the greatest command.
Aborting behaviour is irrecoverable, (i.e.\ $\Abort \Seq c = \Abort$ for any $c$).
\end{description}
For tests, $\cgd{p_1} \refsto \cgd{p_2}$ if $p_1 \supseteq p_2$,
and both $\cpstep{r_1} \refsto \cpstep{r_2}$ and $\cestep{r_1} \refsto \cestep{r_2}$ hold if $r_1 \supseteq r_2$.
Tests/atomic commands are everywhere infeasible for the empty set/relation:
$\cgd{\emptyset} = \cpstep{\emptyset} = \cestep{\emptyset} = \Magic$.
A command is considered \emph{atomic} if it is of the form, $\cpstep{g} \nondet \cestep{r}$, for some relations $g$ and $r$,
that is, it can only make a single transition, 
which may be either a program ($\pstepd$) transition in $g$ or an environment ($\estepd$) transition in $r$.

The command $\Nil$ is the test that always succeeds \refdef{nil}.
The assert command, $\Pre{p}$, aborts if the current state is not in $p$, 
otherwise it terminates immediately \refdef{assert}, where $\compl{p} = \Sigma - p$.
Property \refprop{assert-alt} gives an alternative form for an assert command.
Note that $\Pre{\emptyset} = \Abort$ and $\Pre{\Sigma} = \Nil$.
The command $\cstep{r}$ allows either a program or an environment transition, provided it satisfies $r$ \refdef{cstep}.
The abbreviations $\cpstepd$ and $\cestepd$ allow any program \refdef{cpstepd} or environment \refdef{cestepd} transition, respectively,
and $\cstepd$ allows any transition, program or environment \refdef{cstepd},
where $\universalrel$ is the universal relation between program states \refdef{univ}.
Note the bold fonts for $\Nil$, $\cpstepd$, $\cestepd$ and $\cstepd$.
\\
\begin{minipage}{0.5\textwidth}
\begin{align}
  \Nil & \defs \cgd{\Sigma} \labeldef{nil} \\
  \Pre{p} & \defs \Nil \nondet \cgd{\compl{p}} \Seq \Abort \labeldef{assert} \\
  \Pre{p} & = \cgd{p} \nondet \cgd{\compl{p}} \Seq \Abort \labelprop{assert-alt} \\
  \cstep{r} & \defs \cpstep{r} \nondet \cestep{r} \labeldef{cstep} 
\end{align}
\end{minipage}%
\begin{minipage}{0.5\textwidth}
\begin{align}
  \universalrel & \defs \Sigma \times \Sigma \labeldef{univ} \\
  \cpstepd & \defs \cpstep{\universalrel} \labeldef{cpstepd} \\  
  \cestepd & \defs \cestep{\universalrel} \labeldef{cestepd} \\  
  \cstepd & \defs \cstep{\universalrel} \labeldef{cstepd}  
\end{align}
\end{minipage}
\\[1ex]
Given a set of states $p$ and a relation $r$, 
$r \rres p$ is the relation $r$ restricted so that its range elements are within $p$.
\begin{align}
  r \rres p & \defs \{(\sigma,\sigma') \in r \spot \sigma' \in p\} \labeldef{rres}
\end{align}
Two assertions can be merged \refprop{assert-merge}.
A test after an assertion with the same set of states is redundant \refprop{assert-test}.
If $c$ is refined by $d$ from states in $p_1$ and from states in $p_2$,
then $c$ is refined by $d$ from states in their union \refprop{assert-union},
and that property can be generalised to a set of sets of states $P$, rather than just two sets of states \refprop{assert-Union}.
\begin{align}
  \Pre{p_1} \Seq \Pre{p_2} & = \Pre{p_1 \inter p_2} \labelprop{assert-merge} \\
  \Pre{p} \Seq \cgd{p} & = \Pre{p} \labelprop{assert-test}  \\
  \Pre{p_1 \union p_2} \Seq c & \refsto d & \mbox{if } \Pre{p_1} \Seq c \refsto d \mbox{ and } \Pre{p_2} \Seq c \refsto d \labelprop{assert-union} \\
  \Pre{\textstyle\Union P} \Seq c & \refsto d  & \mbox{ if } \forall p \in P \spot \Pre{p} \Seq c \refsto d  \labelprop{assert-Union}
\end{align}
The inclusion of assertions makes it possible to reason about Hoare triples.

\begin{definitionx}[Hoare-triple]
A command $c$ \emph{establishes} postcondition $p_1$ from initial states satisfying $p$, 
written using the Hoare triple, $\Triple{p}{c}{p_1}$, if and only if,%
\footnote{Note if $c$ does not terminate or aborts, \refprop{triple-form} holds 
because the assertion $\Pre{p_1}$ is not reached,
so this is a weak correctness interpretation of Hoare logic \cite{Wright04}.
For our usages here, $c$ is an expression evaluation, which can neither abort nor fail to terminate.}
\begin{align}
  \Pre{p} \Seq c \refsto c \Seq \Pre{p_1}.  \labelprop{triple-form}
\end{align}
\end{definitionx}
Note the different braces used in the assertion commands and for the pre and post conditions in the Hoare triple.

\subsubsection{Derived commands.}\labelsect{derived}

Finite ($\Fin{c}$) and possibly infinite ($\Om{c}$) iteration of a command $c$ zero or more times
can be defined as the least \refdef{fiter} and greatest \refdef{iter} fixed points, respectively, 
of the monotone function $(\lambda x \spot \Nil \nondet c \Seq x)$.
\\[-2ex]\begin{minipage}{0.5\textwidth}
\begin{align}
  \Fin{c} & \defs \LFP x \spot \Nil \nondet c \Seq x \labeldef{fiter} 
\end{align}
\end{minipage}%
\begin{minipage}{0.5\textwidth}
\begin{align}
  \Om{c} & \defs \GFP x \spot \Nil \nondet c \Seq x \labeldef{iter} 
\end{align}
\end{minipage}\\[2ex]
A program guarantee command, $\guar{g}$, ensures that every program transition satisfies the relation $g$
but places no constraints on environment transitions \refdef{guar}.
A rely command, $\rely{r}$, assumes environment transitions satisfy $r$;
if any environment transition does not, it aborts \refdef{rely},
in the same way that an assertion $\Pre{p}$ aborts if the initial state is not in $p$.
The notation $\compl{r}$ stands for the complement of the relation $r$.
An evolution command \cite{ColletteJones00a},
$\evolve{r}$, combines a rely of $r$ with a guarantee of $r$ \refdef{evolve}.
An invariant command, $\inv{p}$, assumes $p$ holds initially and then 
relies on environment transitions maintaining $p$, and guarantees that program transitions maintain $p$,
i.e.\ it satisfies $\evolve{\postr{p}}$, where $\postr{p}$ is the universal relation restricted so that final states are in $p$ \refdef{postr}.
The command $\Term$ only performs a finite number of program transitions
but does not constrain its environment \refdef{term}.
The command $\Idle$ can perform only a finite number of stuttering (no change) program transitions
but does not constrain its environment \refdef{idle}.
The relation $\id{}$ is the identity relation
and $\id{V}$ is the identity relation on a set of l-values $V$  \refdef{id}.
The command $\Frame{V}{c}$, 
restricts $c$ so that its program transitions can only modify l-values in the set $V$, called its \emph{frame} \refdef{frame}.
Within identity relations 
and frames 
an l-value $u$ stands for the singleton set $\{u\}$.
The command $\opt{r}$ either performs a single program transition satisfying $r$
or it can terminate immediately from states $\sigma$ such that $(\sigma,\sigma) \in r$ \refdef{opt},
that is, $r$ is satisfied by not changing the state.
The atomic specification command, $\atomicrel{r}$, 
achieves $r$ via an atomic program transition 
and allows finite stuttering program transitions (that do not change the state)
and arbitrary environment transitions before and after 
the atomic transition \refdef{atomic-spec};
like $\opt{r}$, the atomic transition satisfying $r$ may be elided for states $\sigma$ for which $(\sigma,\sigma) \in r$. 
\\[-2ex]\begin{minipage}[t]{0.45\textwidth}
\begin{align}
  \guar{r} & \defs \Om{(\cpstep{r} \nondet \cestepd)}  \labeldef{guar} \\
  \rely{r} & \defs \Om{(\cstepd \nondet \cestep{\compl{r}} \Seq \Abort)}  \labeldef{rely} \\
  \evolve{r} & \defs \rely{r} \together \guar{r} \labeldef{evolve} \\
  \postr{p} & \defs \{(\sigma,\sigma') \spot \sigma' \in p\} \labeldef{postr} \\
  \inv{p} & \defs \Pre{p} \Seq \evolve{\postr{p}} \labeldef{inv} \\
  \Term & \defs \Fin{\cstepd} \Seq \Om{\cestepd}  \labeldef{term} 
\end{align}
\end{minipage}%
\begin{minipage}[t]{0.55\textwidth}
\begin{align}
  \Idle & \defs \guar{\id{}} \together \Term  \labeldef{idle} \\
  \id{V} & \defs \{(\sigma,\sigma) \spot \forall v \in V \spot \sigma'\,v = \sigma\,v \} \labeldef{id} \\
  \Frame{V}{c} & \defs \guar{\id{\compl{V}}} \together c \labeldef{frame} \\
  \opt{r} & \defs \cpstep{r} \nondet \cgd{\{\sigma \spot (\sigma,\sigma) \in r\}}  \labeldef{opt} \\
  \atomicrel{r} & \defs \Idle \Seq \opt{r} \Seq \Idle  \labeldef{atomic-spec} 
\end{align}
\end{minipage}\\[1ex]
The postcondition specification command, $\Post{q}$, for $q$ a relation between states,
guarantees to terminate in a final state $\sigma'$ that is related to the initial state $\sigma$ by $q$ \refdef{spec}.
The initial test, $\cgd{\{\sigma\}}$, selects those behaviours that start in state $\sigma$ 
and the final test, $\cgd{\{\sigma' \spot (\sigma,\sigma') \in q\}}$, restricts the behaviours to those satisfying $q$ end-to-end.
The non-deterministic choice accumulates those behaviours for all possible starting states $\sigma$.
\begin{align}
  \Spec{}{}{q} &\defs \Nondet \sigma \spot \cgd{\{\sigma\}} \Seq \Term \Seq \cgd{\{\sigma' \spot (\sigma,\sigma') \in q\}}  \labeldef{spec} 
\end{align}
If a specification command is restricted to achieve a final state satisfying $p$, that restriction may be pulled out as a test \refprop{spec-test}.
The composition of binary relations is written, $r_1 \semi r_2$,
where, $(\sigma,\sigma') \in (r_1 \semi r_2)$ if and only if $(\exists \sigma_1 \spot (\sigma,\sigma_1) \in r_1 \land (\sigma_1,\sigma') \in r_2)$.
A specification command $\Post{r_1 \semi r_2}$ can be split into two specifications, 
where the first establishes the precondition $p$ of the second \refprop{spec-split}.
Guarantees \refprop{guar-distrib-seq} and relies \refprop{rely-distrib-seq} distribute over sequential composition, 
see \cite{2024MeinickeHayesDistributiveLaws}.
\begin{align}
  \Spec{}{}{q \rres p} & = \Spec{}{}{q} \Seq \cgd{p} \labelprop{spec-test} \\
  \Post{r_1 \semi r_2} & \refsto \Post{r_1 \rres p} \Seq \Pre{p} \Seq \Post{r_2} \labelprop{spec-split} \\
  \guar{g} \together c_1 \Seq c_2 & = (\guar{g} \together c_1) \Seq (\guar{g} \together c_2) \labelprop{guar-distrib-seq} \\
  \rely{r} \together c_1 \Seq c_2 & = (\rely{r} \together c_1) \Seq (\rely{r} \together c_2) \labelprop{rely-distrib-seq} 
\end{align}

\ifarxiv\pagebreak[4]\fi
\subsubsection{Expressions.}\labelsect{expressions}

Our treatment of expressions is based on that in \cite{2026MultiReferenceExpressions} and the reader is referred there for more details.
We distinguish between l-values expressions ($\LEx{lve}$) and expressions ($\Ex{e}$).
The syntax of expressions includes 
constants ($\Const{\kappa}$),
dereferencing of l-value expressions ($\Deref{lve}$),
unary operators ($\Unary{\UnaryOp}{e}$),
and
binary operators ($\Binary{e_1}{\BinaryOp}{e_2}$).
L-value expressions consist of either a variable $\Variable{v}$ or an indexed array reference $\Array{lve}{e}$,
in which $\LEx{lve}$ is an l-value expression (thus allowing multi-dimensional arrays)
and $\Ex{e}$ is an index expression.
The notation $\Eval{e}{\sigma}$ stands for the value of expression $\Ex{e}$ in state $\sigma$;
it has the standard inductive definition over the structure of expressions \cite{2026MultiReferenceExpressions}.

In a sequential program the value of an expression is uniquely determined by the state in which it is evaluated,
whereas in a concurrent context its value depends on the sequence of states that occur during its evaluation.
Hence for a given initial state, 
an expression may evaluate to different values depending on any interleaving environment transitions 
between its accesses to variable values.
To allow for such non-determinism, 
we represent the evaluation of an expression $\Ex{e}$ to a value $\Ex{k}$ by the command $\Expr{e}{k}$,
e.g.\ evaluating a pair of expressions $(e_1,e_2)$ to a pair of values $(k_1,k_2)$,
evaluates them in parallel: $\Expr{(e_1,e_2)}{(k_1,k_2)} = \Expr{e_1}{k_1} \parallel \Expr{e_2}{k_2}$.
If $\Ex{e}$ does not evaluate to $\Ex{k}$, $\Expr{e}{k}$ is infeasible.
Because expression evaluation is side-effect free, 
successful evaluation corresponds to the command $\Idle$,
which allows finite stuttering transitions during the evaluation.

Expressions typically occur in a non-deterministic choice over $\Ex{k}$,
for example, the only choice that succeeds for $\Expr{\Const{\kappa}}{k}$ is for $\Ex{k} = \Const{\kappa}$ and 
all other choices for $\Ex{k}$ are infeasible.
The effect when followed by a command, $c\,\Ex{k}$, that is dependent on $\Ex{k}$, is to select $c\,\Const{\kappa}$:
\begin{align*}
  \Nondet \Ex{k} \spot \Expr{\Const{\kappa}}{k} ; c\,\Ex{k} 
     = \Idle \Seq c\,\Const{\kappa} .
\end{align*}

\subsubsection{Programming language commands.}\labelsect{spec-commands}

A conditional command either evaluates the boolean expression $\Ex{b}$ to $\Ex{\True}$ and executes $c$, 
or to $\Ex{\False}$ and executes $d$ \refdef{conditional}.
Because our language is not statically typed, the semantics needs to deal with guards that evaluate to a non-boolean,
in which case conditionals are defined to abort.
\reftheorem{rely-conditional} below includes an assumption \refprop{cond-bool}
that avoids the case when $\Ex{b}$ evaluates to a non-boolean.
A while loop executes $c$ while $\Ex{b}$ evaluates to $\Ex{\True}$ and then terminates if $\Ex{b}$ evaluates to $\Ex{\False}$
or aborts if $\Ex{b}$ evaluates to a non-boolean \refdef{while}.
A greatest fixed point is used because that allows both a finite and an infinite number of iterations,
and hence allows one to address termination of loops. 
For a parameterised command, $f$, that is, $f$ is of type $Value \fun \Command$,
the command, $\Call\,f\,e$, evaluates its argument $e$ to $k$ and then executes $f\,k$; 
the $\Idle$ at the end allows for stuttering transitions to effect the return \refdef{call}.
The parameter passing mechanism is call-by-constant,
that is, the actual parameter $e$ is evaluated to give the value of the formal parameter 
but the formal parameter acts as a constant within the procedure body 
(and hence cannot be modified within the procedure body).
If $e$ is a constant $\kappa$, then $\Call\,f\,\kappa$ reduces to $\Idle \Seq f\,\kappa \Seq \Idle$. 
\begin{align}
  \If b \Then c \Else d \Fi & \defs \Expr{b}{\True} \Seq c \nondet \Expr{b}{\False} \Seq d \nondet \Nondet \{\Expr{b}{k} \Seq \Abort \spot \Ex{k \not\in \bool} \}  \labeldef{conditional} \\
  \While b \Do c \Od & \defs \GFP x .~ \If b \Then c \Seq x \Else \Nil \Fi \labeldef{while} \\
  \Call\,f\,e & \defs \Nondet k \spot \Expr{e}{k} \Seq f \,k \Seq \Idle \labeldef{call}
\end{align}

\subsubsection{Expression evaluation under interference.}\labelsect{expr_interference}

\labelsect{motivation}
Consider the following command,
\begin{align}
  \Pre{p} \Seq \If b \Then \Pre{p_T} \Seq c \Else \Pre{p_F} \Seq d \Fi . \labelprop{if-pt-pf}
\end{align}
In the sequential case \cite{Hoare69a}, 
assuming the initial state satisfies the precondition $p$,
the assertions $p_T$ and $p_F$ in \refprop{if-pt-pf} 
hold if $p_T$ is taken to be $p \inter b$ and $p_F$ is taken to be $p \inter \lnot b$.
However, under interference from concurrent threads, 
the following issues arise:
\begin{enumerate}
\item\labelprop{interference-pre}
the precondition $p$ may be invalidated by interference;
\item\labelprop{interference-true}
if $b$ evaluates to $\True$, it does not follow that $b$ holds at the start of the $\Then$ branch;
and 
\item\labelprop{interference-false}
if $b$ evaluates to $\False$, it does not follow that $\lnot b$ holds at the start of the $\Else$ branch.
\end{enumerate}
To handle the invalidation of the precondition $p$ by interference,
we require that $p$ is stable under the rely condition $r$ \cite{StolenForte92}.
If for a particular example, the precondition $p$ is not stable under $r$,
it can be replaced by a weaker precondition that is stable under $r$.
\begin{definitionx}[stable]
A set of states, $p$, is \emph{stable} under a relation $r$ if, whenever $\sigma \in p$ and $(\sigma,\sigma') \in r$, then $\sigma' \in p$.
\end{definitionx}

For reasoning about guard evaluation under interference we make use of a Hoare triple (\Definition*{Hoare-triple})
of the form, $\Establish{p}{r}{\Expr{b}{k}}{p_1}$,
which can be read as 
``$\Expr{b}{k}$ establishes $p_1$ from initial states satisfying $p$ under interference satisfying $r$''.
For the conditional command in \refprop{if-pt-pf} executing in a context 
in which its environment satisfies a rely condition $r$,
the assertions $p_T$ and $p_F$ in \refprop{if-pt-pf} are valid if the following Hoare triples hold
because these triples correspond to evaluating the guard to $\True$/$\False$ under interference satisfying $r$.\\[-1ex]
\begin{minipage}{0.5\textwidth}
\begin{align}
  \EstablishExpr{p&}{r}{b}{\True}{p_T} \labelprop{establish-pt} 
\end{align}
\end{minipage}%
\begin{minipage}{0.5\textwidth}
\begin{align}
  \EstablishExpr{p&}{r}{b}{\False}{p_F} \labelprop{establish-pf}
\end{align}
\end{minipage}\\[1ex]
These address issues \refprop{interference-true} and \refprop{interference-false} above.
To reason about arbitrary expressions, possibly with multiple references to shared variables,
we make use of a compositional approach in which we establish a postcondition for a complex expression 
by first establishing postconditions for each of its component sub-expressions.
Our approach does not need the syntactic single critical reference restriction \cite{BenAri2005,Andrews1991}
of approaches in which expression evaluation is assumed to be atomic
(see \cite{2026MultiReferenceExpressions} for details).

Because expressions are free from side effects, they refine $\Idle$,
and hence we have the following lemma for introducing an expression evaluation.
See \cite{2026MultiReferenceExpressions} for a proof.
\begin{lemmax}[idle-to-expr-with-post]
For $p$ a set of states, 
$r$ a relation,
$k$ a constant,
$e$ an expression,
and $P\,k$ a set of states,
if 
$\EstablishExpr{p}{r}{e}{k}{P\,\Ex{k}}$,
\begin{align*}
  \rely{r} \together \Pre{p} \Seq \Idle \refsto \rely{r} \together \Expr{e}{k} \Seq \Pre{P\,\Ex{k}} .
\end{align*}
\end{lemmax}

\subsubsection{Introducing a conditional command.}\labelsect{conditional}

To introduce a conditional command,
separate assumptions are required for the case of the guard $\Ex{b}$ evaluating to $\Ex{\True}$ \refprop{cond-true}
and to $\Ex{\False}$ \refprop{cond-false}.
A third assumption \refprop{cond-bool} ensures $\Ex{b}$ evaluates to a boolean value,
which rules out the third aborting alternative in the definition of a conditional \refdef{conditional}.
A fourth assumption \refprop{tolerates-c} requires that $c$ tolerates inference before it starts execution.
See \cite{2026MultiReferenceExpressions} for a proof.
\begin{theoremx}[rely-conditional]
If $\Ex{b}$ is a boolean expression, 
$q$ and $r$ are relations,
$c$ is a command,
$p$, $p_T$ and $p_F$ are sets of states,
and,
\\[-3ex]\begin{minipage}[t]{0.4\textwidth}
\begin{align}
  & \EstablishExpr{p}{r}{b}{\True}{p_T}   \labelprop{cond-true} \\
  & \EstablishExpr{p}{r}{b}{\False}{p_F}   \labelprop{cond-false} 
\end{align}
\end{minipage}%
\begin{minipage}[t]{0.6\textwidth}
\begin{align}
  \forall k \spot \EstablishExpr{p}{r}{b}{k}{\Set{k \in \bool}}   \labelprop{cond-bool} \\
  \rely{r} \together \Pre{p} \Seq c \refsto \rely{r} \together \Pre{p} \Seq \Idle \Seq c \labelprop{tolerates-c}
\end{align}
\end{minipage}\\[1ex]
then
\(
  \rely{r} \together \Pre{p} \Seq c \refsto \If \Ex{b} \Then \rely{r} \together \Pre{p_T} \Seq c \Else \rely{r} \together \Pre{p_F} \Seq c \Fi .
\)
\end{theoremx}

A specification command, $\Post{q}$,  
being refined to a conditional command \refdef{conditional} must tolerate interference satisfying $r$ 
in order to allow for the evaluation of the guard and interference before and after the execution of the conditional \cite{CoJo07}.
\begin{definitionx}[tolerates]
A relation $q$ \emph{tolerates} interference $r$ from precondition $p$ if $p$ is stable under $r$ and both,
\(
  p \dres (r \semi q) \subseteq q  
  ~~\mbox{and}~~
  p \dres (q \semi r) \subseteq q . 
\)
\end{definitionx}
If $q$ tolerates $r$ from $p$, then
$p \dres (\Finrel{r} \semi q \semi \Finrel{r}) \subseteq q$,
where $\Finrel{r}$ is the reflexive, transitive closure of $r$.
That is, $q$ tolerates zero or more $r$ transitions both before and after it,
and thus one can introduce $\Idle$ commands before and after a specification command with postcondition $q$
(see \cite{hayes2021deriving} for a proof of these properties).
\begin{align}
  \rely{r} \together \Pre{p} \Seq \Spec{}{}{q} & = \rely{r} \together \Pre{p} \Seq \Idle \Seq \Spec{}{}{q} \Seq \Idle & \mbox{if $q$ tolerates $r$ from $p$} \labelprop{spec-tolerates}
\end{align}

\subsubsection{Introducing a function call.}\labelsect{call}

A call command first evaluates it actual parameter expression,
which may be subject to interference from parallel threads,
and hence we make use of assumption \refprop{eval-arg},
which establishes the precondition $P\,k$ for the execution of the body of the function (i.e.\ $f\,k$).
To allow for interference both before and after executing the body of the function,
the postcondition relation $q$ must tolerate such interference.
\begin{theoremx}[introduce-call]
If $e$ is an expression,
$p$ and $P\,k$ are set of states, 
$g$, $q$ and $r$ are relations,
where 
$g$ is reflexive
and 
$q$ tolerates $r$ from $p$, 
and
for all $k$, \\[-2ex]
\begin{minipage}{0.4\textwidth}
\begin{align}
   \EstablishExpr{p}{r}{e}{k}{P\,k} \labelprop{eval-arg}
\end{align}
\end{minipage}%
\begin{minipage}{0.6\textwidth}
\begin{align}
   \guar{g} \together \rely{r} \together \Pre{P\,k} \Seq \Post{q} \refsto f\,k
\end{align}
\end{minipage}\\[1ex]
then,
$
  \guar{g} \together \rely{r} \together \Pre{p} \Seq \Post{q} \refsto \Call f\,e~.
$
\ifarxiv{\color{purple}See \reflemy{well-founded-variant}{proofs} for a proof.}\else See \cite{2026RecursionWhileLoops_arxiv} for a proof.\fi
\end{theoremx}
If $p$ is stable under $r$ and the expression $e$ is invariant under interference satisfying $r$,
then $P\,k$ can be taken to be, $p \inter \EqEval{e}{k}$,
where the notation, $\Set{Pr}$, stands for the set of states in which the predicate $Pr$ holds,
where references to program variables within $Pr$ correspond to their value in the state.

\section{Recursion}\labelsect{recursion}

This section develops \reftheorem{recursion-early} that covers introducing recursion.
It uses a variant expression and a well-founded relation $\wfr$ to show termination.
The well-founded induction principle states that, for a well-founded relation $\wfr$
and property $P\,k$ defined on values that are type compatible with $\wfr$, 
\begin{equation}\labelprop{well-founded-induction}
  \left(\forall k \spot (\forall j \spot \WFrelation{j}{k} \implies P\,j) \implies P\,k\right)
    ~~\implies~~ (\forall k \spot P\,k).
\end{equation}
The notation 
$\GtWFEval{e_1}{e_2}$ abbreviates $\Comprehension{}{\WFrelation{\Eval{e_1}{\sigma}}{\Eval{e_2}{\sigma}}}{\sigma}$,
where $\Eval{e}{\sigma}$ is the value of $e$ is state $\sigma$.
The following lemma lifts well-founded induction to reasoning about refinement of commands 
using a parametrised variant expression $z$ that is type compatible with a well-founded relation $\wfr$.
In the following lemma, $s$ and $c$ are parameterised commands, 
that is, of the form $\lambda y \spot body$, for some $body$.
A parameterised command $s$ is refined by another $c$, that is $s \refsto c$, if $\forall y \spot s\,y \refsto c\,y$.
\begin{lemmax}[well-founded-variant]
If
$s$ and $c$ are parameterised commands,
$\wfr$ is a well-founded relation,
$z$ is a parametrised variant expression (type compatible with $\wfr$),
and,
\begin{equation}\labelprop{well-founded-variant-assumption}
  \forall k \spot (\forall y \spot \Pre{\GtWFEval{z\,y}{k}} \Seq s\,y \refsto c\,y) \implies (\forall y \spot \Pre{\EqEval{z\,y}{k}} \Seq s\,y \refsto c\,y)
\end{equation}
then $s \refsto c$.
\ifarxiv{\color{purple}See \reflemy{well-founded-variant}{proofs} for a proof.}\else See \cite{2026RecursionWhileLoops_arxiv} for a proof.\fi
\end{lemmax}

\reftheorem*{recursion-early}
applies \reflem*{well-founded-variant} for $c$ in the form
of the greatest fixed point, $\GFP f$, 
of a monotone function $f$ from parameterised commands to parametrised commands,
that is, the type of $f$ is $(Value \fun \Command) \fun (Value \fun \Command)$,
and hence $\GFP f$ is also a parametrised command.
It is assumed that from states satisfying $p_X$, $f$ does not make any recursive calls
(i.e.\ it only uses its base case(s)),
which is encoded here by $F$ being unconstrained.
The approach corresponds to that of Back and Preoteasa \cite[Theorem 23]{BackPreoteasa2005} for the sequential refinement calculus
but includes the early termination alternative, which is required for derivation of some concurrent programs (see \refsect{ex-early-termin}).
\begin{theoremx}[recursion-early]
If
$p_X$ is a set of states,
 $\wfr$ is a well-founded relation,
$z$ is a parametrised variant expression that is type compatible with $\wfr$, 
$s$ is a parametrised command, 
$f$ is a monotone function from parametrised commands to parametrised commands, 
and both, 
\begin{align}
  \forall F\,y \spot & \Pre{p_X} \Seq s\,y \refsto f\,F\,y \labelprop{well-founded-assumption1} \\
  \forall F\,k \spot & (\forall x \spot \Pre{\GtWFEval{z\,x}{k} \union p_X} \Seq s\,x \refsto F\,x) \implies 
  			     (\forall y \spot \Pre{\EqEval{z\,y}{k}} \Seq s\,y \refsto f\,F\,y) \labelprop{well-founded-assumption2}
\end{align}
then,
$s \refsto \GFP f$.
\ifarxiv{\color{purple}See \reftheoremy{recursion-early}{proofs} for a proof.}\else See \cite{2026RecursionWhileLoops_arxiv} for a proof.\fi
\end{theoremx}
The proviso \refprop{well-founded-assumption1} is used to handle the case in which
$p_X$ holding initially ensures $\GFP f$ does not utilise any recursive calls,
e.g. taking $f$ to be $(\lambda x \spot \If b \Then c \Seq x \Else \Nil \Fi)$, 
the case in which the guard $b$ is guaranteed to evaluate to false if $p_X$ holds.
A special case is if $p_X$ is the empty set, in which case \refprop{well-founded-assumption1} is trivially satisfied
and \refprop{well-founded-assumption2} is simplified.
Note that the rules for recursion are not specific to rely/guarantee concurrency and can be used in other contexts.

\begin{RelatedWork}
Schellhorn et al.~\cite{SchellhornTEPR14} include recursion in their approach, 
in which commands are encoded in a variant of interval temporal logic (RGITL)
and recursive programs are then defined as recursive formulae in the logic.
San{\'a}n et al.\ \cite{Sanan21} allow parameterless procedures and 
make use of a natural number call depth bound to avoid infinite recursion.
Other approaches \cite{CoJo07,Dingel02,PrensaNieto03,DBLP:conf/esop/WickersonDP10,XuRoeverHe97} do not consider recursion,
instead they define the semantics of while loops via an operational semantics,
as do San{\'a}n et al.\ \cite{Sanan21}.
\end{RelatedWork}

\subsection{Example of a parallel recursive search}\labelsect{parallel-search}

The task is to search for a value of $i$ in the range $0 \leq i < N$, such that $P\,i$ holds.
The variable $found$ is set to $\True$ if and only if there exists such a value.
There may be zero or more occurrences of $i$ for which $P\,i$ holds;
any value of $i$ for which $P\,i$ holds is valid.
We use a predicative notation for relations, 
in which a dereferenced l-value expression, $\Deref{lve}$, stands for its value in the pre-state and 
a primed dereferenced l-value expression, $\Deref{lve'}$, stands for its value in the post-state,
for example, 
$\Rel{\Deref{found}' = \Deref{found}} 
= \{ (\sigma,\sigma') \spot \sigma'\,found = \sigma\,found \} .
$
Note that we use lower corners, $\Set{\ldots}$ for sets and upper corners, $\Rel{\ldots}$ for binary relations.
\begin{align}\labeldef{top-level-search}
\begin{array}{l}
  \Rrely{\Deref{found}' = \Deref{found} \land \Deref{i}' = \Deref{i}} \together {} \\
  \Spre{0 \leq N} \Seq 
  \Spec{found,i}{}{\color{blue}\ulcorner((\exists j \spot 0 \leq j < N \land P\,j) \implies \Deref{found}') \land {} \\ \color{blue}~~(\Deref{found}' \implies 0 \leq \Deref{i}' < N \land P(\Deref{i}'))\urcorner}
\end{array}
\end{align}
The function $search$ (specified below) handles searching within a subrange from $lo$ to $hi-1$, inclusive,
which may be empty if $lo \geq hi$.
The evolve command \refdef{evolve} both relies and guarantees that 
once a value $\Deref{i}$ satisfying $P(\Deref{i})$ is found, $\Deref{found}$ remains true.
The invariant \refdef{inv} assumes initially that if $\Deref{found}$ holds so does $P(\Deref{i})$ where $0 \leq \Deref{i} < N$, 
it then relies on any environment transitions maintaining this property and
guarantees that all program transitions maintain the property.
Note that the environment is allowed to change $\Deref{i}$ but only to a value such that $P(\Deref{i})$ holds for $0 \leq \Deref{i} < N$.
\begin{align*}
  search \defs & \lambda (lo,hi) \spot 
   \Revolve{\Deref{found} \implies \Deref{found}'} \together {} \\
  & \Sinv{\Deref{found} \implies 0 \leq \Deref{i} < N \land P(\Deref{i})} \together {} \\
  & \Spre{0 \leq lo \leq hi \leq N} \Seq \Rspec{found,i}{}{(\exists j \spot lo \leq j < hi \land P\,j) \implies \Deref{found}'}
\end{align*}
Note the complete range, from $0$ to $N-1$, is used in the invariant, 
while the post-condition focuses on the range from $lo$ to $hi-1$.
This allows concurrent threads to set $i$ to a value such that $P(\Deref{i})$ holds and set $found$ to be $\True$.
The top level specification \refdef{top-level-search} is implemented by setting found to $\False$ and 
calling $search$ for the range $0$ to $N-1$.
\begin{align*}
  \refdef{top-level-search} \refsto found := \False \Seq \Call search\,(0,N)
\end{align*}
The implementation of $search$ uses divide-and-conquer by splitting the range into two halves and 
recursively searching both halves in parallel.
A parallel search is practical if 
the time complexity of evaluating $P$ is large (relative to the cost of forking threads).
\begin{align*}
  find & \defs \lambda F \spot 
    \begin{array}[t]{l} 
      \lambda (lo,hi) \spot  \Spre{0 \leq lo \leq hi \leq N} \Seq {} \\
      \If \lnot \Deref{found} \land lo+1 = hi \Then ~\Comment{range only contains one element, $lo$} \\
      ~~~~\If P\,lo \Then i := lo \Seq found := \True \Else \Idle \Fi \\
      \Else \If \lnot \Deref{found} \land lo + 2 \leq hi \Then ~\Comment{range has size at least 2} \\
      ~~~~~~~~(\Call F\,(lo,\Mid ) \parallel \Call F\,(\Mid ,hi)) \\
      ~~~~\Else~\Comment{have $\Deref{found}$ already or the range is empty} \\
      ~~~~~~~~\Idle \\
      ~~~~\Fi \\
      \Fi 
   \end{array}
\end{align*}
Function $search$ is implemented recursively as the greatest fixed point of the function $find$,%
\footnote{The programming language Scheme uses the $\kw{letrec}$ notation for such a definition, 
where $fun = \GFP(\lambda F \spot body)$ here corresponds to Scheme's $\mathop{\kw{letrec}} fun = body\rename{F}{fun}$.}
that is, $search \refsto \GFP find$, as can be shown by applying \reftheorem{recursion-early} with $\emptyset$ for $p_X$,
$(\lambda(lo,hi) \spot hi-lo)$ for the variant function $z$, and $search$ for $s$ and $find$ for $f$.
For the theorem, \refprop{well-founded-assumption1} holds trivially because $p_X$ is empty,
and \refprop{well-founded-assumption2} corresponds to the following.
\begin{align}
  \forall F\, k \spot & (\forall l\,h \spot \Pre{\GtWFEval{h - l}{k}} \Seq search(l,h) \refsto F(l,h)) \implies {} \nonumber \\
  			    & (\forall lo\,hi \spot \Pre{\EqEval{hi - lo }{k}} \Seq search(lo,hi) \refsto find\,F\,(lo,hi)) \labelprop{induct}
\end{align}
For all $F$ and $k$, we assume the left side of the implication and show the right side holds.
Our first steps use \reftheorem{rely-conditional} twice to introduce the outer conditional and an inner conditional.
Note that while $\Deref{found}$ is stable under the rely condition and hence can be assumed in the precondition of \refprop{none}, 
$\lnot\Deref{found}$ is not stable and hence can not be assumed in the preconditions of \refprop{one} and \refprop{many}.
\begin{align*}&
  \Spre{hi - lo = k} \Seq search(lo,hi) \refsto \\&
  \If \lnot \Deref{found} \land lo+1 = hi \Then {} \\&
  ~~~~\Spre{lo+1 = hi} \Seq search(lo,hi) \numberthis \labelprop{one} \\&
  \Else \Spre{hi - lo = k} \Seq \\&
  ~~~~\If \lnot \Deref{found} \land lo + 2 \leq hi \Then {} \\&
  ~~~~~~~~\textstyle\Spre{hi - lo = k \land lo < \Mid < hi} \Seq search(lo,hi) \numberthis \labelprop{many} \\&
  ~~~~\Else \\&
  ~~~~~~~~\Spre{\Deref{found} \lor lo = hi} \Seq search(lo,hi) \numberthis \labelprop{none} \\&
  ~~~~\Fi \\&
  \Fi
\end{align*}
For the first alternative \refprop{one}, the range just contains the single value $lo$,
so if $P\,lo$ holds $i$ can be set to $lo$ and $found$ to $\True$,
but it $P\,lo$ does not hold we do nothing (i.e. $\Idle$).
\begin{align*}
  \refprop{one} \refsto \If P\,lo \Then i := lo \Seq found := \True \Else \Idle \Fi
\end{align*}
In the $\Then$ branch, $lo$ must be assigned to $i$ before $\True$ is assigned to $found$ 
in order to preserve the invariant in the intermediate state (i.e.\ just after the assignment to $i$).
In the $\Else$ branch no action is taken, in particular, $found$ is \underline{not} set to $\False$
because it was initialised to $\False$ and if another thread has found an element satisfying $P$,
it will have set $found$ to $\True$ and $found$ should remain $\True$.
The assignment commands in the implementation trivially satisfy the guarantee, $\Rel{\Deref{found} \implies \Deref{found}'}$.

For the second alternative \refprop{many} when $lo+2 \leq hi$, 
a search of the range $(lo,hi)$ needs to establish the postcondition
$(\exists j \spot lo \leq j < hi \land P\,j) \implies \Deref{found}'$,
which can be achieved by two parallel searches that establish, 
$(\exists j \spot lo \leq j < \Mid \land P\,j) \implies \Deref{found}'$ 
and
$(\exists j \spot \Mid \leq j < hi \land P\,j) \implies \Deref{found}'$,
respectively.
The parallel searches use recursive calls on $search$,
which are introduced using \reftheorem{introduce-call}.
\pagebreak[1]
\begin{align*}
  \refprop{many} \refsto & 
  \textstyle \Call\,(\lambda (l,h) \spot \Spre{h - l \wfr k} \Seq search(l,h))\,(lo,\Mid) \parallel \\&
  \textstyle \Call\,(\lambda (l,h) \spot \Spre{h - l \wfr k} \Seq search(l,h))\,(\Mid,hi) \numberthis \labelprop{parallel-search}
\end{align*}
We can then apply the assumption from \refprop{induct} to show,
\begin{align*}
 \refprop{parallel-search} \refsto & \textstyle\Call\,F\,(lo,\Mid) \parallel \Call\,F\,(\Mid,hi)
\end{align*}
Finally, if an element satisfying $P$ has already been found, 
or the range is empty,
the search is satisfied by doing nothing.
\begin{align*}
  \refprop{none} \refsto \Idle
\end{align*}
Combining the above refinement steps gives,
\begin{align}
  \Spre{hi-lo = k} \Seq search(lo,hi) \refsto find\,F\,(lo,hi), 
\end{align}
discharging \refprop{induct},
and hence $search \refsto \GFP\,find$ by \reftheorem{recursion-early}.

\section{Introducing a while loop}\labelsect{loop}

For a while loop we derive a law to establish,
\begin{align}
  \rely{r} \together \Pre{p} \Seq \Post{\Finrel{q} \rres p_F} \refsto \While b \Do c \Od .
\end{align}
The while loop should establish the relation $\Finrel{q}$,
between its initial and final states
and terminate in a state satisfying $p_F$.
The precondition $p$ is also the loop invariant.
To show termination we use a variant expression $z$,
that is required to strictly decrease according to a well-founded relation $\wfr$ on each iteration,
unless an iteration establishes an early termination condition $p_X$ that ensures the loop guard $b$ will evaluate to $\False$
and hence the loop will exit.

The variant expression $z$ may not be increased by interference from concurrent threads \refprop{non-increasing}.
Because the definition of a while loop is based on a conditional command,
the law for introducing a while loop inherits the asumptions that
evaluating the loop guard $\Ex{b}$ to true from states satisfying $p$ under interference satisfying $r$ establishes $p_T$ \refprop{while-true},
and similarly, evaluating $\Ex{b}$ to false establishes $p_F$ \refprop{while-false},
and 
that the guard must evaluate to a boolean \refprop{while-bool}.
If $\Ex{b}$ is guaranteed not to evaluate to $\True$ from states satisfying $p_X \inter p$,
the guard evaluation $\Expr{b}{\True}$ is infeasible and hence establishes $\emptyset$ \refprop{while-infeas}.
From initial states satisfying $p_T$ and that the variant $\Ex{z}$ is less than or equal to $\Ex{k}$,
the body of the loop $c$ must establish $\Finrel{q}$ between its pre and post states,
and its post state must re-establish the loop invariant $p$ and either strictly decrease the variant $\Ex{z}$ or establish $p_X$,
from which the loop is guaranteed to terminate on the next iteration \refprop{while-ref}.
The bound variable $\Ex{k}$ can be thought of as the value of $\Ex{z}$ before guard evaluation;
after guard evaluation only $\Ex{z \wfreq k}$ can be assumed because interference may have decreased $\Ex{z}$,
where $\wfreq$ is the reflexive closure of $\wfr$.
Note that, commonly, $q$ is reflexive and transitive, i.e.\ $q =\Finrel{q}$.
\begin{theoremx}[intro-while]
If
$\Ex{b}$ is a boolean expression, 
$r$ is a relation, 
$p$, $p_T$, $p_F$ and $p_X$ are sets of states, 
$\wfr$ is a transitive, well-founded relation,
$\Ex{z}$ is a variant expression that is type compatible with $\wfr$,
$q$ is a relation, such that $\Finrel{q}$ tolerates $r$ from $p$,
and,
\begin{align}
  & r \subseteq \Rel{z' \wfreq z} \labelprop{non-increasing} \\ 
  & \EstablishExpr{p}{r}{b}{\True}{p_T}  \labelprop{while-true} \\
  & \EstablishExpr{p}{r}{b}{\False}{p_F}  \labelprop{while-false} \\
  \forall k \spot ~& \EstablishExpr{p}{r}{b}{k}{\Set{k \in \bool}}  \labelprop{while-bool} \\
  & \EstablishExpr{p_X \inter p}{r}{b}{\True}{\emptyset} \labelprop{while-infeas} \\
  \forall k \spot ~& \rely{r} \together \Pre{p_T \inter \Set{z \wfreq k}} \Seq \Post{\Finrel{q} \rres ((\Set{z \wfr k} \union p_X) \inter p)} \refsto c \labelprop{while-ref}
\end{align} 
then
\( 
  \rely{r} \together \Pre{p} \Seq \Post{\Finrel{q} \rres p_F} \refsto \While \Ex{b} \Do c \Od .
\)
\ifarxiv{\color{purple}See \reftheoremy{intro-while}{proofs} for a proof.}\else See \cite{2026RecursionWhileLoops_arxiv} for a proof.\fi
\end{theoremx}
A reflexive guarantee distributes into a while loop, that is, a while loop satisfies a guarantee provided its body does.

\subsection{Example using the early termination version of the while loop law}\labelsect{ex-early-termin}
	
To implement an atomic specification command that updates $m$ to the minimum of $m$ and the constant $lo$
under interference that may decrease $m$ but cannot increase $m$,
we can make use of a compare-and-swap instruction, $CAS(m, old, new)$,
that atomically both checks if the current value of the variable $m$ is $old$ and succeeds in updating $m$ to be $new$
but if $m$ is not equal to $old$ it fails and has no effect.
\begin{align*}
  CAS \defs \lambda m\,old\,new\spot \nonumber 
                   \Frame{m}{\atomicrel{(\Deref{m} = old \implies \Deref{m}' = new) \land (\Deref{m} \neq old \implies \Deref{m}' = \Deref{m})}} 
\end{align*}
Because the $CAS$ may fail to update $m$ if the value of $m$ is changed (decreased) 
between its sampling into $old$ and the atomic action of the $CAS$ taking place,
a while loop is required to repeat the $CAS$ until it succeeds in updating $m$ to the minimum of $old$ and $lo$,
each time based on a fresh sample of $\Deref{m}$.%
\footnote{The body of the loop corresponds to $\Let old = \Deref{m} \In \Call\,CAS\,(m, old, min\{lo, old\})$ in other languages.}
\begin{align*}
  & \Rrely{\Deref{m} \geq \Deref{m}'} \together 
     \Frame{m}{\atomicrel{\Rel{\Deref{m}' = min\{lo,\Deref{m}\}}}} \\
  & \refsto \While lo < \Deref{m} \Do 
       \Nondet old \spot \Expr{\Deref{m}}{old} \Seq \Call\,CAS\,(m, old, min\{lo,old\})
     \Od
\end{align*}
Without interference, an obvious loop variant to show the termination of the loop is $\Deref{m}$ 
because (without interference) $\Deref{m}$ is reduced to $lo$ by the $CAS$ 
and the loop terminates after one iteration.
However, interference may at any point decrease the value of $\Deref{m}$.
If that interference decreases $\Deref{m}$ to a value less than $lo$ after the guard evaluation,
the $CAS$ will fail to decrease $\Deref{m}$,
and hence the loop body will fail to decrease the variant $\Deref{m}$.
However, because $\Deref{m}$ has been reduced below $lo$, 
the guard evaluation on the next iteration gives $\False$ and the loop terminates,
which implies we need to utilise the early exit case of the while loop law.
For the application of \reftheorem{intro-while} we take 
$p_X$ as $\Set{\Deref{m} \leq lo}$,
$p_T$ as $\Set{\True}$,
$p_F$ as $\Set{\Deref{m} \leq lo}$,
$p$ as the type invariant $\Set{\Deref{m} \in int \land lo \in int}$,
$q$ as the universal relation $\universalrel$, which is both reflexive and transitive,
and
$z$ as $\Deref{m}$.
Provisos \refprop{non-increasing}, \refprop{while-true}, \refprop{while-false} and \refprop{while-infeas} hold trivially,
and \refprop{while-bool} follows from the type invariant $p$.
Finally the body of the loop satisfies \refprop{while-ref} 
because if the $CAS$ succeeds, $p_X$ is stably established and guarantees both the postcondition and termination of the loop,
whereas if the $CAS$ fails, interference must have decreased $\Deref{m}$, that is, it has reduced the variant.

\subsection{Fair termination}\labelsect{fairness}

The command $\Fair$ disallows an infinite contiguous sequence of environment transitions \cite{FM2018fairness}.
\begin{align}
  \Fair \defs \Om{(\Fin{\estepd} \Seq \pstepd)} \Seq \Fin{\estepd} \labeldef{fair}
\end{align}
Fair execution of a command $c$ can be represented as $c \together \Fair$,
which rules out an infinite contiguous sequence of environment transitions when executing $c$, unless $c$ aborts.
The command $\Term$ allows only a finite number of program transitions
but does not rule out an infinite contiguous sequence of environment transitions,
however, $\Term \together \Fair = \Fin{\cstepd}$ and hence allows only a finite number of transitions.
Thus if a command $c$ refines $\Term$, then $c \together \Fair$ only performs a finite number of transitions, i.e. 
\begin{align}
  c \together \Fair \refines \Term \together \Fair = \Fin{\cstepd} .
\end{align}
Hence it is sufficient to show that a command $c$ refines $\Term$ 
in order to show fair execution of $c$ only makes a finite number of transitions.
The reader is referred to \cite{FM2018fairness} for a more thorough treatment of fairness in our theory.

\begin{RelatedWork}
Dingel \cite{Dingel02} handles fairness by defining the trace semantics of his parallel operator to only allow fair interleaving.
Because the semantics of Schellhorn et al.\ \cite{SchellhornTEPR14} explicitly alternates between a program transition and an environment transition
(corresponding to any number of our environment transitions),
it implicitly corresponds to fair execution.
Our approach allows one to decouple fair execution and allows a simpler definition of the parallel operator \cite{FM2018fairness}.
\end{RelatedWork}

\section{Conclusion}\labelsect{conclusion}

The main contributions of this work are to provide
general
laws for reasoning about recursion and while loops in the context of interference from concurrent threads,
including termination arguments based on a variant and a well-founded relation.
Our laws cover early termination because this is useful in the context of concurrency,
and in this context (unlike the sequential context) they do not follow from a version of a law that does not handle the early termination case. 
Our approach handles reasoning about fair termination in the context of concurrency,
that is, termination provided the thread is executed fairly \cite{FM2018fairness}.

We do not assume either expression evaluations or assignment commands are executed atomically
but employ a fine-grained semantics for expressions to justify a set of inference rules
that allow compositional reasoning about expressions, including guards \cite{2026MultiReferenceExpressions}.
The fine-grained theory has been used to develop laws for 
assignment and conditional commands in \cite{2026MultiReferenceExpressions}
and for procedure calls, recursion and while loops in the current paper.
We only support call-by-constant parameter passing 
where formal parameters act as constants and hence may not be modified within the procedure body
but because we allow an l-value to be passed (by constant) as a parameter, 
we effectively support call-by-reference parameters.
More general call-by-value parameter passing 
where the formal parameter acts as a local variable that can be modified within the procedure body,
and call-by-result parameters and combined call-by value-result parameters
could be supported in future work by combining the theories developed here 
with our theories for handling local variables \cite{FormaliSE_localisation}.

Our concurrency theory has been formalised within
the Isabelle/\-HOL theorem prover \cite{IsabelleHOL}, and the results
presented in this paper have been verified in that theory.
\ifreview\marginpar{\color{purple}For reviewing purposes, a pdf of the Isabelle/HOL theories for expressions and commands are included after the Appendix.}\fi

\paragraph{Acknowledgements.}
Our collaboration has been supported by the 
Australian Research Council  
under their Discovery Program Grant 
DP190102142.
Hayes and Meinicke were also supported by funding from the 
Department of Defence, administered through the Advanced Strategic Capabilities Accelerator
grant \emph{Verifying Concurrent Data Structures for Trustworthy Systems}.
Jones was supported by
Leverhulme Trust
RPG-2019-020
and would like to acknowledge the stimulus of the  {\em  Big Specification} programme at the Cambridge {\em Isaac Newton Institute}.

\bibliography{ms}

\ifarxiv

\newpage
\appendix
\renewcommand\theHsection{\thesection.appendix}

\section{Proofs}\labelapp{proofs}

\begin{theoremy}[introduce-call]
If $e$ is an expression,
$p$ and $P\,k$ are set of states, 
$g$, $q$ and $r$ are relations,
where 
$g$ is reflexive
and 
$q$ tolerates $r$ from $p$, 
and
for all $k$, \\[-2ex]
\begin{minipage}{0.4\textwidth}
\begin{align}
   \EstablishExpr{p}{r}{e}{k}{P\,k} \labelprop{eval-arg1}
\end{align}
\end{minipage}%
\begin{minipage}{0.6\textwidth}
\begin{align}
   \guar{g} \together \rely{r} \together \Pre{P\,k} \Seq \Post{q} \refsto f\,k \labelprop{intro-f}
\end{align}
\end{minipage}\\[1ex]
then,
$
  \guar{g} \together \rely{r} \together \Pre{p} \Seq \Post{q} \refsto \Call f\,e~.
$
\end{theoremy}

\begin{proof}
For all $k$, we have,
\begin{align*}&
  \guar{g} \together \rely{r} \together \Pre{p} \Seq \Post{q}
 \Equals*[by \refprop{spec-tolerates} as $q$ tolerates $r$ from $p$]
  \guar{g} \together \rely{r} \together \Pre{p} \Seq \Idle \Seq \Post{q} \Seq \Idle
 \Refsto*[by \reflem{idle-to-expr-with-post} using \refprop{eval-arg1}]
  \guar{g} \together \rely{r} \together \Expr{e}{k}  \Seq \Pre{P\,k} \Seq \Post{q} \Seq \Idle
 \Refsto*[distribute guar \refprop{guar-distrib-seq} and rely \refprop{rely-distrib-seq}; remove from $\Expr{e}{k}$ and $\Idle$ as $g$ is reflexive]
  \Expr{e}{k} \Seq (\guar{g} \together \rely{r} \together \Pre{P\,k} \Seq \Post{q}) \Seq \Idle
 \Refsto*[by \refprop{intro-f}]
  \Expr{e}{k} \Seq f\,k \Seq \Idle
\end{align*}
and hence by \refdef{call}, $\guar{g} \together \rely{r} \together \Pre{p} \Seq \Post{q} \refsto \Nondet k \spot  \Expr{e}{k} \Seq f\,k \Seq \Idle = \Call\,f\,e$.
\end{proof}

\begin{lemmay}[well-founded-variant]
If 
$s$ and $c$ are parametrised commands,
$\wfr$ is a well-founded relation,
$z$ is a parametrised variant expression that is type compatible with $\wfr$,
and,
\begin{equation}\labelprop{well-founded-variant-assumption-y}
  \forall k \spot (\forall y \spot \Pre{\GtWFEval{z\,y}{k}} \Seq s\,y \refsto c\,y) \implies (\forall y \spot \Pre{\EqEval{z\,y}{k}} \Seq s\,y \refsto c\,y)
\end{equation}
then $s \refsto c$.
\end{lemmay}

\begin{proof}
By well-founded induction \refprop{well-founded-induction}, 
with $P\,k$ as $(\forall y\spot \Pre{\Set{z\,y = k}} \Seq s\,y \refsto c\,y)$,
the universal quantification $\forall k \spot P\,k$, holds if for all $k$, 
\begin{align}
  (\forall j \spot j \wfr k \implies (\forall y \spot \Pre{\Set{z\,y = j}} \Seq s\,y \refsto c\,y)) \implies (\forall y \spot \Pre{\Set{z\,y = k}} \Seq s\,y \refsto c\,y),
\end{align}
which holds as follows.
\begin{align*}&
  \forall j \spot j \wfr k \implies (\forall y \spot \Pre{\Set{z\,y = j}} \Seq s\,y \refsto c\,y)
 \Entails*[by \refprop{assert-Union} with its $P$ as $\{ \Set{z\,y = j} \mid j \spot j \wfr k \}$]
  \forall y \spot \Pre{\textstyle\Union \{ \Set{z\,y = j} \mid j \spot j \wfr k \}} \Seq s\,y \refsto c\,y
 \Equiv*[as $\Union \{ \Set{z\,y = j} \mid j \spot j \wfr k \} = \Set{z\,y \wfr k}$]
  \forall y \spot \Pre{\Set{z\,y \wfr k}} \Seq s\,y \refsto c\,y
 \Entails*[by assumption \refprop{well-founded-variant-assumption-y}]
  \forall y \spot \Pre{\Set{z\,y = k}} \Seq s\,y \refsto c\,y
\end{align*}
and hence by well-founded induction \refprop{well-founded-induction}, 
$\forall k \spot \forall y \spot \Pre{\Set{z\,y = k}} \Seq s\,y \refsto c\,y$,
and hence by \refprop{assert-Union},
$\forall y \spot \Pre{\Union k \spot \Set{z\,y = k}} \Seq s\,y \refsto c\,y$,
and hence as $(\Union k \spot \Set{z\,y = k}) = \Sigma$,
we can deduce that $\forall y \spot s\,y \refsto c\,y$, that is, $s \refsto c$.
\end{proof}

\begin{theoremy}[recursion-early]
If
$p_X$ is a set of states,
$\wfr$ is a well-founded relation,
$z$ is a parametrised variant expression that is type compatible with $\wfr$, 
$s$ is a parametrised command, 
$f$ is a monotone function from parametrised commands to parametrised commands, 
and both,
\begin{align}
  \forall F\,y \spot & \Pre{p_X} \Seq s\,y \refsto f\,F\,y \labelprop{well-founded-assumption1-y} \\
  \forall F\,k \spot & (\forall x \spot \Pre{\GtWFEval{z\,x}{k} \union p_X} \Seq s\,x \refsto F\,x) \implies 
  			      (\forall y \spot \Pre{\EqEval{z\,y}{k}} \Seq s\,y \refsto f\,F\,y) \labelprop{well-founded-assumption2-y}
\end{align}
then,
$s \refsto \GFP\,f$.
\end{theoremy}

\begin{proof}
If we instantiate $F$ in \refprop{well-founded-assumption1-y} with $\GFP\,f$ 
and then simplify because $\GFP\,f$ is a fixed point of $f$, that is,
\begin{equation}
  f(\GFP\,f) = \GFP\,f  \labelprop{unfold-gfp}
\end{equation}
we obtain,
\begin{align}
  \forall y \spot & \Pre{p_X} \Seq s\,y \refsto f\,(\GFP\,f)\,y = (\GFP\,f)\,y \labelprop{px-gfp} 
\end{align}
The proof uses \reflem{well-founded-variant} with $\GFP\,f$ for $c$,
and hence one must show,
\begin{align}
  \forall k \spot (\forall y \spot \Pre{\GtWFEval{z\,y}{k}} \Seq s\,y \refsto (\GFP\,f)\,y) \implies 
  			(\forall y \spot \Pre{\EqEval{z\,y}{k}} \Seq s\,y \refsto (\GFP\,f)\,y)
\end{align}
For any $k$, we start from the left side of the implication and show the right side.
\begin{align*}&
  \forall y \spot \Pre{\GtWFEval{z\,y}{k}} \Seq s\,y \refsto (\GFP\,f)\,y
 \Implies*[using \refprop{assert-union} to combine above with \refprop{px-gfp}]
  \forall y \spot \Pre{\GtWFEval{z\,y}{k} \union p_X} \Seq s\,y \refsto (\GFP\,f)\,y
 \Implies*[using \refprop{well-founded-assumption2-y} with $\GFP\,f$ for $F$]
  \forall y \spot \Pre{\EqEval{z\,y}{k}} \Seq s\,y \refsto f\,(\GFP\,f)\,y
 \Implies*[folding the fixed point by \refprop{unfold-gfp}]
  \forall y \spot \Pre{\EqEval{z\,y}{k}} \Seq s\,y \refsto (\GFP\,f)\,y
  \qedhere
\end{align*}
\end{proof}

\begin{theoremy}[intro-while]
If
$\Ex{b}$ is a boolean expression, 
$r$ is a relation, 
$p$, $p_T$, $p_F$ and $p_X$ are sets of states, 
$\wfr$ is a transitive, well-founded relation,
$z$  is a variant expression that is type compatible with $\wfr$,
$q$ is a relation, such that $\Finrel{q}$ tolerates $r$ from $p$,
and,
\begin{align}
  & r \subseteq \Rel{z' \wfreq z} \labelprop{non-increasing-y} \\ 
  & \EstablishExpr{p}{r}{b}{\True}{p_T}  \labelprop{while-true-y} \\
  & \EstablishExpr{p}{r}{b}{\False}{p_F}  \labelprop{while-false-y} \\
  \forall k \spot ~& \EstablishExpr{p}{r}{b}{k}{\Set{k \in \bool}}  \labelprop{while-bool-y} \\
  & \EstablishExpr{p_X \inter p}{r}{b}{\True}{\emptyset} \labelprop{while-infeas-y} \\
  \forall k \spot ~& \rely{r} \together \Pre{p_T \inter \Set{z \wfreq k}} \Seq \Post{\Finrel{q} \rres ((\Set{z \wfr k} \union p_X) \inter p)} \refsto c \labelprop{while-ref-y}
\end{align} 
then
\( 
  \rely{r} \together \Pre{p} \Seq \Post{\Finrel{q} \rres p_F} \refsto \While \Ex{b} \Do c \Od .
\)
\end{theoremy}

\begin{proof}
Because a while loop is defined as a greatest fixed point of the monotone function 
$(\lambda x \spot \If b \Then c \Seq x \Else \Nil \Fi)$,
we make use of \reftheorem{recursion-early} to prove the law,
except we do not require the parameter $y$ here, so quantifications over $y$ and its uses as a parameter are elided.
The provisos for \reftheorem{recursion-early}, 
with $\rely{r} \together \Pre{p} \Seq \Post{\Finrel{q} \rres p_F} $ for $s$
and $(\lambda x \spot \If b \Then c \Seq x \Else \Nil \Fi)$ for $f$,
are the following.
\begin{align}
  \forall F \spot \Pre{p_X} \Seq (\rely{r} \together \Pre{p} \Seq \Post{\Finrel{q} \rres p_F})  \refsto \If b \Then c \Seq F \Else \Nil \Fi \labelprop{px-infeasible} \\
  \begin{array}{ll}
    \forall F\,k \spot &\Pre{\Set{z \wfr k} \union p_X} \Seq (\rely{r} \together \Pre{p} \Seq \Post{\Finrel{q} \rres p_F}) \refsto F \\
       & \implies \Spre{z = k} \Seq (\rely{r} \together \Pre{p} \Seq \Post{\Finrel{q} \rres p_F}) \refsto \If b \Then c \Seq F \Else \Nil \Fi 
  \end{array}\labelprop{refines}
\end{align}
The left side of \refprop{px-infeasible} is equivalent to $\rely{r} \together \Pre{p_X \inter p} \Seq \Post{\Finrel{q} \rres p_F}$
and hence to show the refinement \refprop{px-infeasible} one can use \reftheorem{rely-conditional} with the proof obligations:
\begin{align}
  \Triple{p_X \inter p}{&\rely{r} \together \Expr{b}{\True}}{\emptyset} \labelprop{px-b-true} \\
  \Triple{p_X \inter p}{&\rely{r} \together \Expr{b}{\False}}{p_F} \labelprop{px-b-false} \\
  \forall k \spot \Triple{p_X \inter p}{&\rely{r} \together \Expr{b}{k}}{\Set{k \in \bool}} \labelprop{px-b-bool} \\
  \rely{r} \together \Pre{p_X \inter p} \Seq \Post{\Finrel{q} \rres p_F} & \refsto \rely{r} \together \Pre{p_X \inter p} \Seq \Idle \Seq \Post{\Finrel{q} \rres p_F} \labelprop{px-tolerates}
\end{align}
and then show the refinement of the $\Then$ and $\Else$ branches of the conditional,
\begin{align}
  \rely{r} \together \Pre{\emptyset} \Seq \Post{\Finrel{q} \rres p_F} \refsto &~c \Seq F \labelprop{px-c1} \\
  \rely{r} \together \Pre{p_F} \Seq \Post{\Finrel{q} \rres p_F} \refsto &~\Nil \labelprop{px-c2}
\end{align}
Property \refprop{px-b-true} is just assumption \refprop{while-infeas-y},
\refprop{px-b-false} follows from assumption \refprop{while-false-y},
similarly, \refprop{px-b-bool} follows from \refprop{while-bool-y},
and \refprop{px-tolerates} holds by \refprop{spec-tolerates} because $\Finrel{q}$ tolerates $r$ from $p$ by assumption.
Property \refprop{px-c1} holds because $\Pre{\emptyset} = \Abort$ 
and hence the left side reduces to $\Abort$, which is refined by any other command.
Property \refprop{px-c2} holds because relation $\Finrel{q}$ is reflexive 
and hence $\Post{\Finrel{q}} \refsto \Post{\id{}} \refsto \Nil$
and hence 
$\rely{r} \together \Pre{p_F} \Seq \Post{\Finrel{q} \rres p_F} 
\refsto \Pre{p_F} \Seq \Post{\Finrel{q}} \Seq \cgd{p_F} 
\refsto \Pre{p_F} \Seq \cgd{p_F}
= \Pre{p_F}
\refsto \Nil$ by \refprop{assert-test}.

To show \refprop{refines}, we assume for all $F$ and $k$ that the left side of the implication holds 
and show the right side holds but weaken the precondition on the left of the refinement from $\Set{z = k}$ to $\Set{z \wfreq k}$,
which after merging the preconditions \refprop{assert-merge} requires one to show,
\begin{align}
  \rely{r} \together \Pre{\Set{z \wfreq k}  \inter p} \Seq \Post{\Finrel{q} \rres p_F} \refsto \If b \Then c \Seq F \Else \Nil \Fi 
\end{align}
which can be accomplished using \reftheorem{rely-conditional} with the proof obligations:
\begin{align}
  \EstablishExpr{\Set{z \wfreq k} \inter p&}{r}{b}{\True}{p_T \inter \Set{z \wfreq k}}   \labelprop{ref-true} \\
  \EstablishExpr{\Set{z \wfreq k} \inter p&}{r}{b}{\False}{p_F}   \labelprop{ref-false} \\
  \forall k \spot \EstablishExpr{\Set{z \wfreq k} \inter p&}{r}{b}{k}{\Set{k \in \bool}}   \labelprop{ref-bool} \\
  \rely{r} \together \Pre{\Set{z \wfreq k} \inter p} \Seq \Post{\Finrel{q} \rres p_F} & \refsto \rely{r} \together \Pre{\Set{z \wfreq k} \inter p} \Seq \Idle \Seq \Post{\Finrel{q} \rres p_F} \labelprop{ref-tolerates}
\end{align}
and then show the refinement of the $\Then$ and $\Else$ branches of the conditional,
\begin{align}
  \rely{r} \together \Pre{p_T \inter \Set{z \wfreq k}} \Seq \Post{\Finrel{q} \rres p_F} \refsto & ~c \Seq F \labelprop{ref-c1} \\
  \rely{r} \together \Pre{p_F} \Seq \Post{\Finrel{q} \rres p_F} \refsto &~\Nil \labelprop{ref-c2}	
\end{align}
Property \refprop{ref-true} holds by assumptions \refprop{while-true-y} and \refprop{non-increasing-y},
where \refprop{non-increasing-y} ensures that $\Set{z \wfreq k}$ is stable under the rely $r$ and hence holds after the guard evaluation.
Properties  \refprop{ref-false} and \refprop{ref-bool} follow from 
assumptions \refprop{while-false-y} and \refprop{while-bool-y}, respectively.
Property \refprop{ref-tolerates} holds by \refprop{spec-tolerates} because $\Finrel{q}$ tolerates $r$ from $p$ by assumption.
Property \refprop{ref-c2} is the same as \refprop{px-c2}, which was shown above.
Property \refprop{ref-c1} holds as follows.
\begin{align*}&
  \rely{r} \together \Pre{p_T \inter \Set{z \wfreq k}} \Seq \Post{\Finrel{q} \rres p_F} 
 \Equals*[applying \refprop{spec-test} to pull out the final test and as $\Finrel{q}$ is transitive, $\Finrel{q} = \Finrel{q} \semi \Finrel{q}$]
  \rely{r} \together \Pre{p_T \inter \Set{z \wfreq k}} \Seq \Post{\Finrel{q} \semi \Finrel{q}} \Seq \cgd{p_F} 
 \Refsto*[using \refprop{spec-split} to split $\Post{\Finrel{q} \semi \Finrel{q}}$ and introduce assertion $(\Set{z \wfr k} \union p_X) \inter p$]
  \rely{r} \together \begin{array}[t]{l}
  			    \Pre{p_T \inter \Set{z \wfreq k}} \Seq \Post{\Finrel{q} \rres ((\Set{z \wfr k} \union p_X) \inter p)} \Seq {} \\
  			    \Pre{(\Set{z \wfr k} \union p_X) \inter p} \Seq \Post{\Finrel{q}} \Seq \cgd{p_F} 
			   \end{array}
 \Equals*[distribute the rely over the sequential composition \refprop{rely-distrib-seq}]
  (\rely{r} \together \Pre{p_T \inter \Set{z \wfreq k}} \Seq \Post{\Finrel{q} \rres ((\Set{z \wfr k} \union p_X) \inter p)}) \Seq \\&
  (\rely{r} \together \Pre{(\Set{z \wfr k} \union p_X) \inter p} \Seq \Post{\Finrel{q}} \Seq \cgd{p_F}) 
 \Refsto*[apply assumption \refprop{while-ref-y} to introduce $c$]
  c \Seq (\rely{r} \together \Pre{(\Set{z \wfr k} \union p_X) \inter p} \Seq \Post{\Finrel{q}} \Seq \cgd{p_F}) 
 \Equals*[split assertion \refprop{assert-merge} and pull out the first assertion; apply \refprop{spec-test} in reverse]
  c \Seq \Pre{\Set{z \wfr k} \union p_X} \Seq (\rely{r} \together \Pre{p}  \Seq \Post{\Finrel{q} \rres p_F}) 
 \Refsto*[applying the left side assumption in \refprop{refines}]
  c \Seq F
 \qedhere
\end{align*}
\end{proof}

\fi

\ifreview
\newpage
\fi

\end{document}